\documentclass[aps,prl,twocolumn,groupedaddress]{revtex4}
\usepackage{graphicx}
\usepackage{calc}
\usepackage{pifont}
\usepackage{amssymb}
\usepackage{epsfig}
\usepackage{amssymb}
\usepackage{amsmath}
\usepackage{color}
\DeclareMathAlphabet{\mathitb}{OT1}{cmr}{bx}{sl}
\begin{document}

\renewcommand{\thefootnote}{\fnsymbol{footnote}}
\title{Mechanism of Enhanced Rectification in Unimolecular Borromean Ring Devices}
\author{Gavin D. Scott$^1$}
\email{gscott@physics.ucla.edu}
\author{Kelly S. Chichak$^2$}
\author{Andrea J. Peters$^2$}
\author{Stuart J. Cantrill$^2$}
\author{J. Fraser Stoddart$^2$}
\author{H. W. Jiang$^1$}

\affiliation{
$^1$Department of Physics and Astronomy, UCLA, Los Angeles, CA 90095\\
$^2$Department of Chemistry and Biochemistry, UCLA, Los Angeles, CA 90095}

\date{\today}

\begin{abstract}

We have studied charge transport through individual Borromean Ring
complexes, both with and without anchor groups, in gated double
barrier tunneling junctions (DBTJs) formed using the electrical
breakjunction technique on gold nanowires.  While common single
molecule device characteristics can be observed with either form of
the Borromean Rings, the complexes with anchor groups show strong
rectification of conduction in a relatively high percentage of
samples.  We present our data along with a simple model underlining
the mechanism by which the arrangement and composition of the weakly
bonding anchor groups attached to the electroactive element may
promote a device configuration resulting in rectification.

\end{abstract}

\maketitle

The field of molecular electronics has grown dramatically in recent
years as efforts to explore new limits of device scaling have
surged.  Diodes represent one of the most basic and essential
components of modern electronics, so it is no surprise that a
molecular version of the diode has been the focus of many research
efforts.  There is more than one process by which asymmetrical
conduction, i.e. rectification, can occur in metal-molecule-metal
assemblies.  The most heavily investigated mechanism of molecular
rectification is based on the Aviram-Ratner
proposal\cite{Aviram1974}, which utilizes a molecule with two
insulated molecular orbitals, whose probability amplitudes are
asymmetrically oriented with respect to the equilibrium Fermi level
of the source and drain electrodes.\cite{Ellenbogen2000,
Krzeminski2001}  Another process occurs if the molecule, or the
molecular orbital through which current must pass, is situated
asymmetrically within the metal-molecule-metal
junction.\cite{Kornilovitch2002}  In a breakjunction device, for
example, the asymmetrical position of the molecule between the
electrodes corresponds to asymmetrical tunneling barriers through
which charge must be transported.  Because the majority of the
applied voltage drops across the larger insulating barrier, the
conduction conditions for charge transport through a level are
achieved at very different voltages for positive and negative bias.
Our experimental observations of rectification through Borromean
Ring complexes are based on this mechanism.

The devices we fabricated are composed of a source and a drain
electrode plus a conducting substrate that can be used as a back
gate.  We employed the electrical breakjunction technique first
described by Park et al.\cite{HPark1999}  The molecular compound
used as the active element was a Borromean Ring (BR) complex (Figure
1a-c), the synthesis of which was described
recently.\cite{Chichak2004}  The control group used bare BR
complexes with no anchor groups (or ``legs'') and the experimental
group used BR complexes possessing six anchor groups.

The accumulated data indicates that, like many molecules situated
within a breakjunction gap, BR complexes are capable of producing a
range of transport properties (including rectification), each of
which can be anticipated with a certain frequency of
occurrence.\cite{Yu2004}  The complexes with legs did not result in
a greater percentage of devices containing a single molecule.
However, the single molecule devices fabricated with the
experimental group exhibited device characteristics with more
frequent and marked asymmetries, indicating that the property of
conduction rectification can be promoted by the addition of anchor
groups on this molecular compound.  Treating these unimolecular
devices as DBTJs, we may invoke the model described by Kornilovitch
et al.,\cite{Kornilovitch2002}, which indicates that I-V asymmetries
result from the specific arrangement of the metal-molecule-metal
junction.  We speculate that the general affinity for the complexes
of the experimental group to result in a configuration within the
breakjunction gap producing such enhanced rectification
characteristics is due largely to the arrangement and bonding mode
of the six anchor groups that protrude from the BR complex.

The BR complex consists of three interlocking molecular rings,
composed of macrocyclic organic ligands interacting with six
zinc(II) ions (Fig. 1a-c).  The complex was constructed by multiple
cooperative self-assembly processes from eighteen original
components - six endo-tridentate ligands, six exo-bidentate ligands,
and six transition metal ions.  These complexes are relatively large
($\sim$ 2.3 nm in diameter), making them well suited for use in
conjunction with the breakjunction method of fabricating single
molecule devices.  These complexes are also roughly symmetric in
shape.  With this relative lack of anisotropy we hope to alleviate
inconsistencies that may arise when measuring conductance through
its various axes.

\begin{figure}[!h]
\begin{center}
\includegraphics [scale = 0.34]{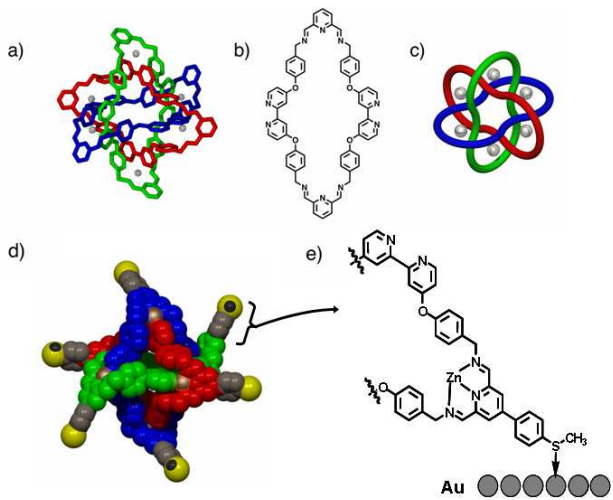}
\end{center}
\vspace{-6mm} \caption{(Color online)  a) The X-ray crystal
structure of the zinc(II)-containing Borromean Ring complex.  b) The
constitution of the macrocyclic ligand, which is repeated three
times over in the Borromean Ring complex.  c) Graphical
representation of the Borromean Ring complex.  d) Graphical
representation of the complex with anchor groups.  e) Chemical
structure of anchor group showing dative bonding to a gold atom.}
\label{figure1}
\end{figure}

The six ``legs'' on the experimental group of BRs are methyl
thioether substituents.  The methyl thioethers are coupled to the
Schiff base precursor, namely the dialdehyde, by employing a
palladium catalyzed Suzuki cross-coupling reaction between the
4-bromo-2, 6-diformylpyridine and 4-methylthiophenyl boronic
acid.\cite{Chichak2005}  The anchor groups extend from the body of
the BR complex in an octahedrally displaced orientation (Fig. 1d).
Among similar research efforts, the prevalent mode of attaching
anchor groups to gold electrodes is through covalent
bonding.\cite{Reed1997,Reichert2002,Ashwell2006}  The covalently
bound mode is formed from the free thiol, i.e. R-SH.  The sulfur
atom is negatively charged and the gold is formally positive
producing a strong chemical bond.  The methyl thioether legs affixed
to the BR complex attach to gold in the dative mode of binding (Fig.
1e).  An electron pair of the sulfur atom is donated to gold into an
empty orbital.  Both gold and sulfur are neutral, hence the bond is
not as strong as the covalent mode for gold-sulfur.

Fabrication of the DBTJ structures begins with the preparation of
arrays of devices on an n+ Si substrate with a 140 \AA\ SiO$_2$
insulating layer.  1.5 nm of Ti and 13 nm of Au are evaporated onto
nanowire patterns defined by electron beam lithography.
Bowtie-shaped constriction patterns are produced with minimum widths
of less than 100 nm.  Additional gold pads for wire-bonding are
defined using UV lithography, followed by the deposition of 10 nm Ti
and 200 nm Au.  The array of samples is cleaned, and a dilute
solution of the BR complex (1 mg/2 ml) is deposited onto the
devices.\cite{Yu2003}  The devices are then cooled to 4.2 K, after
which the electrical breakjunction technique is used to produce a
nanometer-size gap, located at the narrowest portion of the
nanowires.

Due to the probabilistic nature of a given sample's specific
configuration, a statistical approach is taken to ascertain the
device characteristics.  Our experimental data was accumulated from
the results of approximately 450 samples.  After the gap forming
process is complete, differential conductance as a function of
source-drain voltage is measured for every device.  The I-V plots
that result typically fit into one of four categories (detailed by
Yu and Natelson\cite{Yu2004}), each of which represents a general
electrode-molecule configuration within the assembly.

\vspace{0mm}
\begin{figure}[!]
\begin{center}
\includegraphics [scale = 0.28]{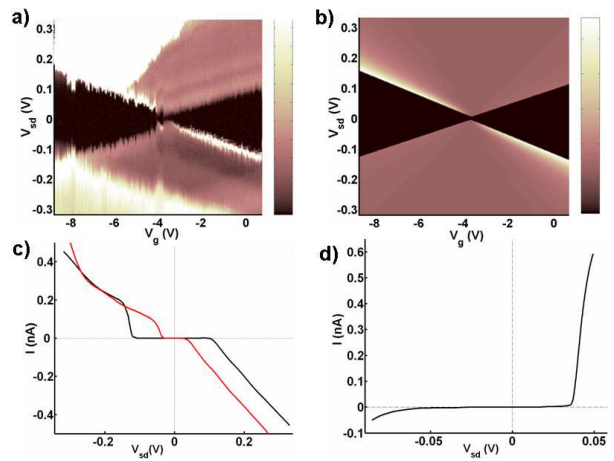}
\end{center}
\vspace{-6mm} \caption{(Color online)  a) Stability diagram of a
single molecule transistor utilizing a BR complex.  b) Simulation of
a gated DBTJ using $C_d \sim C_s$.  For (a) and (b) black represents
zero conductance and white represents the max conductance ($10^{-7}$
nS).  A change in charge state of the BR complex occurs at the
critical point, $V_g$ = $V_C$, where the conductance gap goes to
zero ($\sim$ -3.6 V).  c) Two I-V curves from the stability diagram
in Fig 2a, obtained by integrating along the traces at $V_g$ = 0.7 V
(black) and $V_g$ = -2.3 V (red).  Both I-V plots have rectification
ratios of order unity.  d) I-V plot of device with rectification
ratio of $\sim$ 200.} \label{Plots1c}
\end{figure}

A Breakjunction device with a moderately sized gap in which a single
molecule or complex is present (i.e. a metal-molecule-metal
configuration) will have an I-V plot exhibiting a blockaded region
around $V_{sd} = 0$ with strong nonlinearity outside this region.
Conduction occurs as electrons sequentially tunnel from one
electrode onto the active element and then onto the second
electrode, as established in the seminal works on the
subject.\cite{HPark2000,JPark2002}  This can also be confirmed by
computer simulation.  Fig 2a shows an experimentally obtained
stability diagram (or colormap) measuring differential conductance
($\partial{I}/\partial{V}$) as a function of both source-drain bias
($V_{sd}$) and gate voltage ($V_{g}$).  Treating the tunneling
barriers as though comprised of a capacitive and a resistive
element, our computer model of electrons sequentially tunneling
through a DBTJ can project a theoretical stability diagram with
nearly identical ground state properties as those found in the
experimental data set (Fig 2b).  The positive and negative slopes of
the blockaded region (tunneling thresholds) are determined by the
ratios $C_g/C_d$ and $C_g/C_s$, respectively\cite{Chamapgne2005}
($C_g$, $C_d$, and $C_s$ are the capacitances between the
molecule-gate, molecule-drain, and molecule-source electrodes).  The
relative slopes of the tunneling thresholds in Fig 2a indicate that
the BR complex is almost equally coupled to the source and drain
electrodes, and non-rectifying I-V plots result (Fig 2c).  We
contend that highly asymmetric I-V plots, as in Fig 2d, result when
the electroactive element has a stronger coupling to either the
source or drain electrode.

A device qualified as a ``rectifier'' if it possessed a
rectification ratio greater than 20.  Of the BR samples without
anchor groups, 37$\%$ were confirmed to consist of a
metal-molecule-metal configuration.  Of these ``working'' single
molecule devices, 3$\%$ had rectification ratios $>$ 20. For the BR
samples with anchor groups there was a 35$\%$ chance of a device
resulting in a metal-molecule-metal configuration, but of these
devices over 15$\%$ had rectification ratios $>$ 20.  Furthermore,
only one device among the control group had a rectification ratio
$>$ 50.  Among the devices fabricated using the BR complex with
methyl thioether anchor groups, half of the rectifying samples had
rectification ratios $>$ 100.

The single molecule assembly consists of five parts:  a source and
drain electrode, the central complex through which electrons must
pass, and two insulating barriers (Fig. 3a,b).  The tunneling
barriers are established by the physical separation between the
complex and an electrode and/or the strength of the bond between an
electrode and one or more anchor groups of the BR complex.  The
rectification properties of the assembly are primarily dictated by
three parameters:  the energy difference between the equilibrium
Fermi level of the electrodes and the nearest molecular orbital
level, $\Delta$ (HOMO level\cite{Ghosh2002, Kergueris1999}), and the
coupling strengths of the molecule-source and molecule-drain
tunneling barriers, $\Gamma_s$ and $\Gamma_d$.  The voltage drops
across the barriers are proportional to their respective coupling
strengths.  The ratio of the voltage drops on the opposing barriers,
$\eta$, is $\sim$ $\Gamma_s$$/$$\Gamma_d$.

\vspace{0mm}
\begin{figure}[!ht]
\begin{center}
\includegraphics [scale = 0.29]{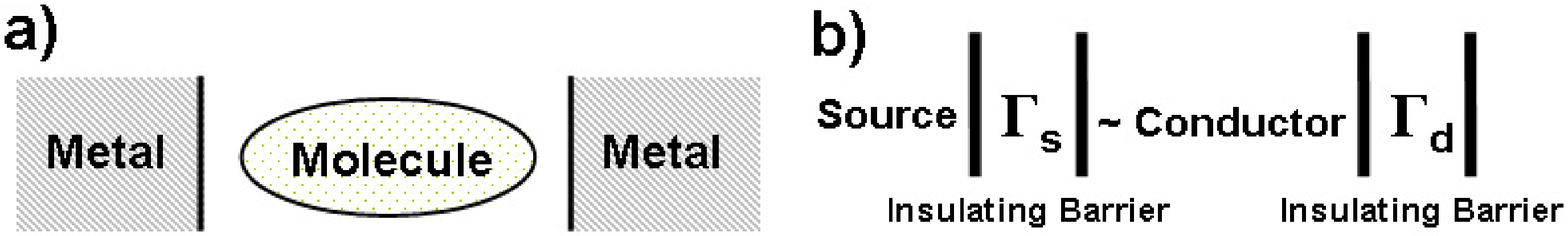}
\includegraphics [scale = 0.33]{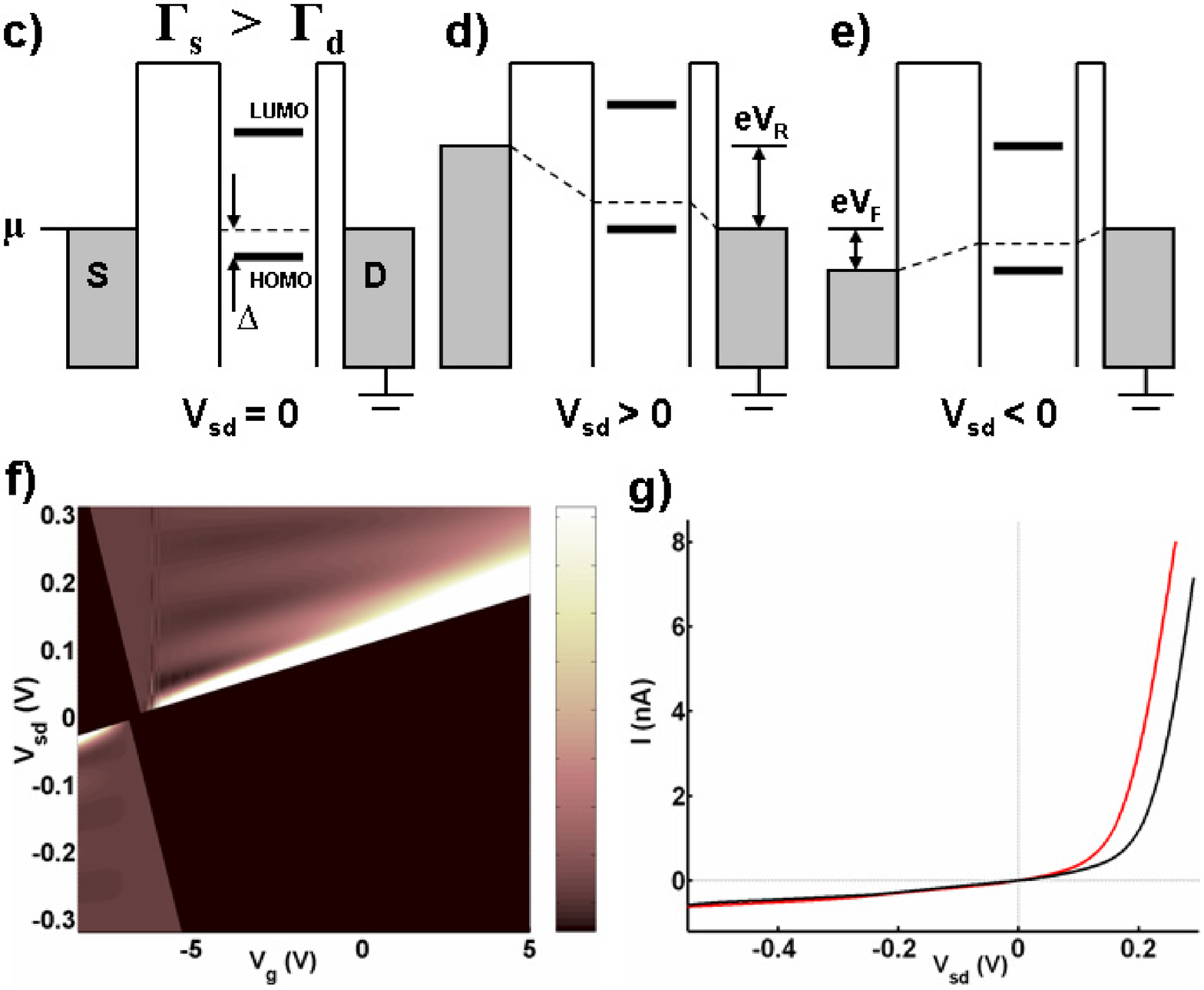}
\end{center}
\vspace{-6mm} \caption{(Color online)  a) Basic molecule-electrode
configuration.  b) The five components of a single molecule
assembly.  c) Energy level diagram of electrodes and BR complex for
$V_{sd} = 0$.  d) For positive bias (forward voltage direction), the
current will increase when the HOMO level of the complex aligns with
the Fermi level of the source electrode.  e) For negative bias
(reverse voltage direction), the current will increase when the HOMO
level of the complex aligns with the Fermi level of the drain
electrode.  A voltage must be applied such that the energy of the
HOMO level goes up by $\Delta$.  The majority of the voltage drop
occurs over the drain-molecule tunneling barrier.  This causes $V_F
> V_R$.  f) Simulation using same parameters as Fig 2b, except
$C_d > C_s$.  $V_g$ offset due to background charge.  Max
$\partial{I}/\partial{V}$ = $10^{-6}$ nS  g) Measured I-V traces
with rectification ration of $\sim$ 25, taken at $V_g$ = 0 V (red)
and $V_g$ = 3 V (black).  The rectification is less pronounced
compared to Fig 2d, but the current density is an order of magnitude
greater before breakdown.} \label{Plots2b}
\end{figure}

The mechanism of rectification follows from Kornilovitch et
al.\cite{Kornilovitch2002}  The ``turn on'' voltages for the forward
and reverse directions are
\begin{equation}
V_R = (1 + \eta)\displaystyle{\frac{\Delta}{e}}, \hspace{1cm} V_F =
\displaystyle\left({\frac{1 +
\eta}{\eta}}\right)\displaystyle{\frac{\Delta}{e}}
\end{equation}
The ratio of the forward and reverse turn-on voltages is $V_R/V_F =
\eta$.  For any $\eta$ greater than 1, there is a voltage window
$V_F < |V| < V_R$ within which there is significant conduction in
the forward direction, but little or no conduction in the reverse
direction (Fig 3c-d).  As the difference in coupling strength
between opposing sides of the complex grows, $\eta$ becomes larger
and hence, the difference between the two threshold voltages
increases.

Fig 3f shows another simulation using nearly identical parameters as
those used to produce Fig 2b with the exception that $C_d$ has been
increased and $C_s$ has been reduced to represent a shift in
position (or coupling strength) of the complex toward the drain
electrode.  The reduced slope of the tunneling threshold for
$V_{sd}, V_{g} > 0$, indicates that a relatively large change in
gate bias causes little change in the position of the forward ``turn
on'' voltage, $V_F$.  Unlike the two curves in Fig 2c, the I-V
measurements in Fig 3g undergo a more modest transformation as the
gate bias is shifted, which is in agreement with our simulation
($V_C$ and $V_R$ cannot be observed due to the onset of significant
instability leading to electrostatic breakdown).  Hence
rectification can derive from the molecule-electrode configuration,
and like typical single molecule transistor characteristics, results
from a single particle sequential tunneling mechanism.

\begin{figure}[!h]
\begin{center}
\includegraphics [scale = 0.3]{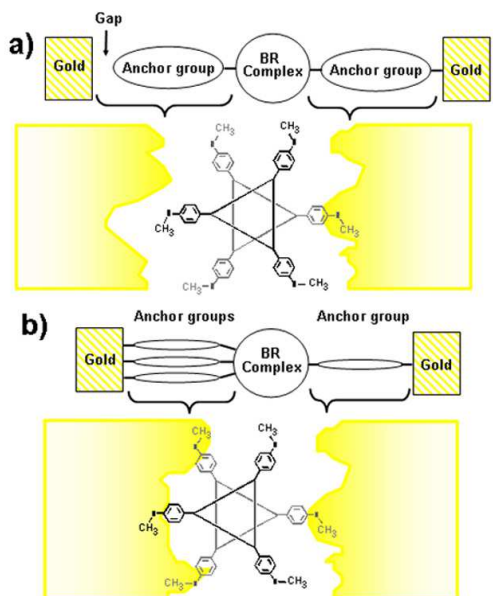}
\end{center}
\vspace{-6mm} \caption{(Color online)  Two possible
electrode-molecule configurations resulting in conduction
rectification.  a) The complex is chemically bonded to only one
electrode.  b) The complex is attached to opposing electrodes
through an unequal number of bonds.  The asymmetry has no
preferential orientation (i.e. either tunneling barrier can be the
larger).  The black and gray triangles represent the two facial
planes of the octahedral complex.} \label{figure4}
\end{figure}

Compromising factors, however, limit the effectiveness of these
rectifiers.  If the bulk of the BR complex is too close to one
electrode, or it is too strongly coupled, the conducting level
widens such that the transmission probability will be significant
for all energies, and there will be little difference between the
currents in the forward and reverse directions.  Conversely, the
larger of the two tunneling barriers controls the overall amplitude
of the current as well as the size of the rectification window.  As
the width of this barrier is increased, the rectification ratio
increases linearly while the current decays exponentially.

Compared to the control complexes with no legs, the Borromean Ring
complexes with methyl thioethers anchor groups resulted in an equal
percentage of devices with an electrode-(single molecule)-electrode
configuration, but five times the percentage of devices with strong
rectification.  Thus the addition of anchor groups on the BR complex
promotes rectification, which we ascertain to result from
molecule-electrode configurational asymmetries.  The means by which
the anchor groups force this configurational asymmetry is a matter
of speculation.  The methyl thioethers legs have a weak affinity for
gold, providing some level of stability, but the chances of both
source and drain ends forming chemical bonds is reduced since the
mode of binding is less favored compared to the covalent bonding
mode.  This weaker dative bonding mode implies that more assemblies
may have one side bonded and the other side free (Fig. 4a), leading
to strongly asymmetric tunneling barriers.

Rectification may also be encouraged via the geometry of the anchor
groups.  The radially protruding anchor groups may act like bumper
guards to stop the complex from positioning itself so close to an
electrode that the energy levels are overly broadened.  Also,
because there are six anchor groups, three on each of the two
tripodal facial planes, configurations may occur in which the source
and drain electrodes are connected to the complex through an unequal
number of bonds, again resulting in asymmetrical tunneling barriers
(Fig 4b).

Conduction asymmetries due to unequal coupling to the electrodes or
to an asymmetric central molecular unit have been observed in the
past but, current rectification ratios in these studies were of
order unity, making them unsuitable for practical
applications.\cite{Reichert2002}  The addition of anchor groups on
the BR complex play a pivotal role in the resulting
molecule-electrode configuration, specifically in promoting
configurations with strongly asymmetric tunneling barriers.  We
postulate that the geometry and bonding mechanism of the anchor
groups on the BR complexes precipitate this outcome.  Affecting the
preferential configuration of a molecule within a breakjunction
represents a significant step toward a more complete understanding
of physics on the molecular level and ultimately extends our ability
to exert further control over such nanoscale systems.

This work is supported by the Defense MicroElectronics Activity
(DMEA 90-02-2-0217) and by the National Science Foundation
(CHE0317170).


\vspace{-6mm}
\begin{thebibliography}{99}

\bibitem{Aviram1974}A. Aviram and M. A. Ratner, Chem. Phys. Lett., {\bf 29}, 277 (1974).

\bibitem{Ellenbogen2000}J. C. Ellenbogen and J. C. Love, Proc. IEEE, {\bf 88}, 386 (2000).

\bibitem{Krzeminski2001}C. Krzeminski, C. Delerue, G. Allan, D. Vuillaume, and R. M. Metzger, Phys. Rev. B, {\bf 64}, 085405 (2001).

\bibitem{Kornilovitch2002}P. E. Kornilovitch, A. M. Bratkovsky, and R. S. Williams, Phys. Rev. B, {\bf 66}, 165436 (2002).

\bibitem{HPark1999}H. Park, A. K. L. Lim, A. P. Alivisatos, J. Park, and P. L. McEuen, Appl. Phys. Lett. \textbf{75}, 301 (1999).

\bibitem{Chichak2004}K. S. Chichak, S. J. Cantrill, A. R. Pease, S. H. Chiu, G. W. V. Cave, J. L. Atwood, J. F. Stoddart, Science \textbf{304}, 1308 (2004).

\bibitem{Yu2004}L. H. Yu and D. Natelson, Nanotechnology \textbf{15}, S517 (2004).

\bibitem{Chichak2005}K. S. Chichak, A. J. Peters, S. J. Cantrill, and J. F. Stoddart, J. Org. Chem., {\bf 70}, 7956 (2005).

\bibitem{Reed1997}M. A. Reed, C. Zhou, C. J. Muller, T. P. Burgin, and J. M. Tour, Science \textbf{278}, 252 (1997).

\bibitem{Reichert2002}J. Reichert, R. Ochs, D. Beckmann, H. W. Weber, M. Mayor, and H. v L\"{o}hneysen, Phys. Rev. Lett., \textbf{88}, 176804 (2002).

\bibitem{Ashwell2006}G. J. Ashwell and A. Chwialkowska, Chem. commun., 1404 (2006).

\bibitem{Yu2003}L. H. Yu and D. Natelson, Nano Lett. \textbf{4}, 79 (2003).

\bibitem{HPark2000}H. Park, J. Park, A. K. L. Lim,
E. H. Anderson, A. P. Alivisatos, and P. L. McEuen, Nature
\textbf{407}, 57 (2000).

\bibitem{JPark2002}J. Park, A. N. Pasupathy, J. I.
Goldsmith, C. Chang, Y. Yaish, J. R. Petta, M. Rinkoski, J. P.
Sethna, H. D. Abr\'{u}na, P. L. McEuen, and D. C. Ralph, Nature
\textbf{417}, 722 (2002).

\bibitem{Chamapgne2005}A. Champagne, A. N. Pasupathy, and D. C. Ralph, Nano Lett. \textbf{5}, 305 (2005).

\bibitem{Ghosh2002}A. W. Ghosh, F. Zahid, S. Datta, and R. R. Birge, Chem. Phys. \textbf{281}, 225 (2002).

\bibitem{Kergueris1999}C. Kergueris, J. P. Bourgoin, S. Palacin, D. Esteve, C. Urbina, M. Magoga, and C. Joachim, Phys. Rev. B \textbf{59}, 12505 (1999).

\end{thebibliography}
\end{document}